\documentstyle[aps,epsfig,12pt]{revtex}

\tightenlines

\begin{document}
ADP-04-07/T588

JLAB-THY-04-3
\begin{center}
\vspace*{2cm} {\Large {\bf New treatment of the
chiral {\em\bf SU(3)} quark mean field model}}\\[0pt]
\vspace*{1cm} P. Wang$^a$, D. B. Leinweber$^a$, A. W. Thomas$^{a,b}$ and A. G. Williams$^a$ \\[0pt]
\vspace*{0.2cm} {\it $^a$Special Research Center for the Subatomic Structure of
Matter (CSSM) and Department of Physics, University of Adelaide 5005, Australia } \\[0pt]
\vspace*{0.2cm} {\it $^b$Jefferson Laboratory, 12000 Jefferson Ave., Newport News, VA 23606 USA}\\[0pt]
\end{center}

\begin{abstract}
We perform a study of infinite hadronic matter, finite nuclei and hypernuclei
with an improved method of calculating the
effective baryon mass. A detailed study of the predictions of the model is
made in comparison with the available data and the level of agreement
is generally very good. Comparison with an earlier treatment shows
relatively minor differences at or below normal nuclear matter density,
while at high density the improved calculation is quite different.
In particular, we find no phase
transition corresponding to chiral symmetry restoration in high density
nuclear matter.

\end{abstract}

\bigskip

\leftline{PACS number(s): 21.65.+f; 21.10-k; 21.80+a; 12.39.-x}
\bigskip
\leftline{{\bf Keywords: Hadronic Matter, Finite Nuclei, Hypernuclei, Effective Mass,}}

{\bf ~~~~~~~~~~ Chiral Symmetry, Quark Mean Field}

\section{Introduction}

The complex action of QCD at finite chemical potential makes it
difficult to study finite-density QCD properties directly from first
principle lattice calculations. There are many
phenomenological models based on hadron degrees of freedom,
such as the Walecka model \cite{Walecka}, the Zimanyi-Mozkovski model
\cite{Zimanyi} and the nonlinear $\sigma-\omega$ model \cite{Bodmer},
as well as models which explicitly include quark degrees of
freedom, for example, the quark meson coupling model \cite{Guichon},
the cloudy bag model \cite{Thomas}, the quark mean field model \cite{Toki}\
and the NJL model \cite{Bentz}. With the
development of these phenomenological models, the interactions
between hadrons become more and more complex. The symmetries of
QCD can be used to determine largely how the hadrons should
interact with each other. With this in mind, the chiral $SU(2)$
$\times$ $SU(2)$ effective quark model was proposed. It
has been used widely to investigate nuclear matter and finite
nuclei at both zero and finite temperature \cite{Heide}-\cite{Qian}.

In the last twenty years,
exploring systems with strangeness, especially with a large
strangeness fraction, has attracted a lot of interest. Therefore,
a model which includes strange quarks or hyperons is needed. Papazoglou
$et$ $al$. \cite{Papazoglou1,Papazoglou2} proposed a
chiral $SU(3)$ model which extended
the $SU(2)$ chiral symmetry to $SU(3)$$\times$$SU(3)$. Recently,
a chiral $SU(3)$ quark mean field model based on
quark degrees of freedom was proposed by Wang $et$ $al$. \cite{Wang3,Wang4}.
In this model, quarks are confined in the baryons by an effective
potential. The quark-meson interaction and meson self-interaction
are based on $SU(3)$ chiral symmetry. Through the mechanism of
spontaneous symmetry breaking the resulting constituent quarks
and mesons (except for the pseudoscalars) obtain masses. The
introduction of an explicit symmetry breaking term in the meson
self-interaction generates the masses of the pseudoscalar mesons
which satisfy the relevant PCAC relations. The explicit symmetry
breaking term in the quark-meson interaction gives reasonable
hyperon potentials in hadronic matter. This chiral $SU(3)$ quark
mean field model has been applied to investigate nuclear matter \cite{Wang2},
strange hadronic matter \cite{Wang3}, finite nuclei, hypernuclei \cite{Wang4},
and quark matter \cite{Wang5}. By and large the results are in reasonable
agreement with existing experimental data.

A surprising result, which is quite different from most other
models is that at some critical density the effective baryon mass
drops to zero \cite{Wang6}. There is a first order phase transition
around this critical density where the physical quantities change
discontinuously. We will see later that this behavior is primarily
a consequence of the nonlinear ansatz for the effective baryon mass which
is related to the subtraction of the centre of mass (c.m.) motion.
In Ref.~\cite{Guichon2}, the authors provided an exact solution of
the effect of c.m.\ motion and found that it was only
very weakly dependent on the
external field strength for the densities of interest. As a result,
the c.m.\ correction can be included in the zero point energy.
In this paper, we will use this alternative definition of the effective
baryon mass in medium to reexamine the properties of
hadronic matter.

The paper is organized as follows. The model is introduced in section II.
In section III, we study infinite hadronic matter, finite nuclei,
and hypernuclei with the improved treatment of the centre of mass
correction to the baryon mass in medium. The numerical
results are presented in section IV and section V is the summary.

\bigskip
\bigskip

\section{The model}

Our considerations are based on the chiral $SU(3)$ quark mean field model
(for details see Refs.~\cite{Wang3,Wang4}), which contains
quarks and mesons as basic degrees of freedom.
In the chiral limit, the quark field $q$ can be split into left and
right-handed parts $q_{L}$ and $q_{R}$: $q\,=\,q_{L}\,+\,q_{R}$.
Under $SU(3)$$_{L}\times$ $SU(3)$$_{R}$ they transform as
\begin{equation}
q_{L} \rightarrow q_{L}^{\prime }\,=\,L\,q_{L},~~~~~
q_{R} \rightarrow q_{R}^{\prime }\,=\,R\,q_{R}\,.
\end{equation}
The spin-0 mesons are written in the compact form
\begin{equation}
M(M^{+})=\Sigma \pm i\Pi =\frac{1}{\sqrt{2}}\sum_{a=0}^{8}
\left( s^{a}\pm i p ^{a}\right) \lambda ^{a},
\end{equation}
where $s^{a}$ and $p ^{a}$ are the nonets of scalar and pseudoscalar
mesons, respectively, $\lambda ^{a}(a=1,...,8)$ are the Gell-Mann
matrices, and $\lambda ^{0}=\sqrt{\frac{2}{3}}\,I$. The alternatives,
plus and minus signs correspond to $M$ and $M^{+}$. Under chiral
$SU(3)$ transformations, $M$ and $M^{+}$ transform as
$M\rightarrow M^{\prime }=LMR^{+}$ and $M^{+}\rightarrow
M^{+^{\prime }}=RM^{+}L^{+}$. The spin-1
mesons are arranged in a similar way as
\begin{equation}
l_{\mu }(r_{\mu })=\frac{1}{2}\left( V_{\mu }\pm A_{\mu }\right)
= \frac{1}{2\sqrt{2}}\sum_{a=0}^{8}\left( v_{\mu }^{a}\pm a_{\mu }^{a}
\right) \lambda^{a}
\end{equation}
with the transformation properties:
$l_{\mu}\rightarrow l_{\mu }^{\prime }=Ll_{\mu }L^{+}$,
$r_{\mu}\rightarrow r_{\mu }^{\prime }=Rr_{\mu }R^{+}$.
The matrices $\Sigma$, $\Pi$, $V_{\mu }$ and $A_{\mu }$ can be
written in a form where the physical states are explicit.
For the scalar and vector nonets, we have the expressions
\begin{eqnarray}
\Sigma = \frac1{\sqrt{2}}\sum_{a=0}^8 s^a \, \lambda^a=\left(
\begin{array}{lcr}
\frac1{\sqrt{2}}\left(\sigma+a_0^0\right) & a_0^+ & K^{*+} \\
a_0^- & \frac1{\sqrt{2}}\left(\sigma-a_0^0\right) & K^{*0} \\
K^{*-} & \bar{K}^{*0} & \zeta
\end{array}
\right),
\end{eqnarray}
\begin{eqnarray}
V_\mu = \frac1{\sqrt{2}}\sum_{a=0}^8 v_\mu^a \, \lambda^a=\left(
\begin{array}{lcr}
\frac1{\sqrt{2}}\left(\omega_\mu+\rho_\mu^0\right)
& \rho_\mu^+ & K_\mu^{*+}\\
\rho_\mu^- & \frac1{\sqrt{2}}\left(\omega_\mu-\rho_\mu^0\right)
& K_\mu^{*0}\\
K_\mu^{*-} & \bar{K}_\mu^{*0} & \phi_\mu
\end{array}
\right).
\end{eqnarray}
Pseudoscalar and pseudovector nonet mesons can be written in
a similar fashion.

The total effective Lagrangian is written:
\begin{eqnarray}
{\cal L}_{{\rm eff}} \, = \, {\cal L}_{q0} \, + \, {\cal L}_{qM}
\, + \,
{\cal L}_{\Sigma\Sigma} \,+\, {\cal L}_{VV} \,+\, {\cal L}_{\chi SB}\,
+ \, {\cal L}_{\Delta m_s} \, + \, {\cal L}_{h}, + \, {\cal L}_{c},
\end{eqnarray}
where ${\cal L}_{q0} =\bar q \, i\gamma^\mu \partial_\mu \, q$ is the
free part for massless quarks. The quark-meson interaction
${\cal L}_{qM}$ can be written in a chiral $SU(3)$ invariant way as
\begin{eqnarray}
{\cal L}_{qM}=g_s\left(\bar{\Psi}_LM\Psi_R+\bar{\Psi}_RM^+\Psi_L\right)
-g_v\left(\bar{\Psi}_L\gamma^\mu l_\mu\Psi_L+\bar{\Psi}_R\gamma^\mu
r_\mu\Psi_R\right)~~~~~~~~~~~~~~~~~~~~~~~  \nonumber \\
=\frac{g_s}{\sqrt{2}}\bar{\Psi}\left(\sum_{a=0}^8 s_a\lambda_a
+ i \gamma^5 \sum_{a=0}^8 p_a\lambda_a
\right)\Psi -\frac{g_v}{2\sqrt{2}}
\bar{\Psi}\left( \gamma^\mu \sum_{a=0}^8
 v_\mu^a\lambda_a
- \gamma^\mu\gamma^5 \sum_{a=0}^8
a_\mu^a\lambda_a\right)\Psi.
\end{eqnarray}
In the mean field approximation, the chiral-invariant scalar meson
${\cal L}_{\Sigma\Sigma}$ and vector meson ${\cal L}_{VV}$
self-interaction terms are written as~\cite{Wang3,Wang4}
\begin{eqnarray}
{\cal L}_{\Sigma\Sigma} &=& -\frac{1}{2} \, k_0\chi^2
\left(\sigma^2+\zeta^2\right)+k_1 \left(\sigma^2+\zeta^2\right)^2
+k_2\left(\frac{\sigma^4}2 +\zeta^4\right)+k_3\chi\sigma^2\zeta
\nonumber \\ \label{scalar}
&&-k_4\chi^4-\frac14\chi^4 \ln \frac{\chi^4}{\chi_0^4} +
\frac{\delta}
3\chi^4 \ln \frac{\sigma^2\zeta}{\sigma_0^2\zeta_0}, \\
{\cal L}_{VV}&=&\frac{1}{2} \, \frac{\chi^2}{\chi_0^2} \left(
m_\omega^2\omega^2+m_\rho^2\rho^2+m_\phi^2\phi^2\right)+g_4
\left(\omega^4+6\omega^2\rho^2+\rho^4+2\phi^4\right), \label{vector}
\end{eqnarray}
where $\delta = 6/33$; $\sigma_0$, $\zeta_0$ and $\chi_0$ are the
vacuum expectation values of the corresponding mean fields
$\sigma$, $\zeta$ and $\chi$.

From the quark-meson interaction, the coupling constants between
scalar mesons, vector mesons and quarks have the following
relations:
\begin{eqnarray}
\frac{g_s}{\sqrt{2}}
&=& g_{a_0}^u = -g_{a_0}^d = g_\sigma^u = g_\sigma^d = \ldots =
\frac{1}{\sqrt{2}}g_\zeta^s,
~~~~~g_{a_0}^s = g_\sigma^s = g_\zeta^u = g_\zeta^d = 0 \, ,\\
\frac{g_v}{2\sqrt{2}}
&=& g_{\rho^0}^u = -g_{\rho^0}^d = g_\omega^u = g_\omega^d = \ldots =
\frac{1}{\sqrt{2}}g_\phi^s,
~~~~~g_\omega^s = g_{\rho^0}^s = g_\phi^u = g_\phi^d = 0 .
\end{eqnarray}
Note, the values of $\sigma_0$, $\zeta_0$ and $\chi_0$ are determined
from a minimization of the thermodynamic potential.
On the other hand,
the parameters $\sigma_0$ and $\zeta_0$ are constrained by the
spontaneous breaking of chiral symmetry and are expressed by
the pion ($F_\pi$ = 93~MeV) and the kaon ($F_K$ = 115~MeV)
leptonic decay constants as:
\begin{eqnarray}\label{sigma_0}
\sigma_0 = - F_\pi    \hspace*{1cm} \hspace*{1cm}
\zeta_0  = \frac{1}{\sqrt{2}} ( F_\pi - 2 F_K)
\end{eqnarray}

The Lagrangian ${\cal L}_{\chi SB}$ generates the
nonvanishing masses of pseudoscalar mesons
\begin{equation}\label{L_SB}
{\cal L}_{\chi SB}=\frac{\chi^2}{\chi_0^2}\left[m_\pi^2F_\pi\sigma +
\left(
\sqrt{2} \, m_K^2F_K-\frac{m_\pi^2}{\sqrt{2}} F_\pi\right)\zeta\right],
\end{equation}
leading to a nonvanishing divergence of the axial currents which in
turn satisfy the partial conserved axial-vector current (PCAC)
relations for $\pi$ and $K$ mesons. Pseudoscalar,
scalar mesons and also the dilaton field $\chi$ obtain mass terms by
spontaneous breaking of chiral symmetry in the Lagrangian
(\ref{scalar}). The masses of $u$, $d$ and $s$ quarks are generated by
the vacuum expectation values of the two scalar mesons $\sigma$ and
$\zeta$. To obtain the correct constituent mass of the strange quark,
an additional mass term has to be added:
\begin{eqnarray}
{\cal L}_{\Delta m_s} = - \Delta m_s \bar q S q
\end{eqnarray}
where $S \, = \, \frac{1}{3} \, \left(I - \lambda_8\sqrt{3}\right) =
{\rm diag}(0,0,1)$ is the strangeness quark matrix. Based on these
mechanisms, the quark constituent masses are finally given by
\begin{eqnarray}
m_u=m_d=-\frac{g_s}{\sqrt{2}}\sigma_0
\hspace*{.5cm} \mbox{and} \hspace*{.5cm}
m_s=-g_s \zeta_0 + \Delta m_s.
\end{eqnarray}
The parameters $g_s = 4.76$ and $\Delta m_s = 29$~MeV are chosen to yield
the constituent quark masses $m_q=313$~MeV and $m_s=490$~MeV.
In order to obtain reasonable hyperon potentials in hadronic matter, we
include an additional coupling between strange quarks and the scalar
mesons $\sigma$ and $\zeta$~\cite{Wang3}. This term is expressed as
\begin{eqnarray}
{\cal L}_h \,=\, (h_1 \, \sigma \, + \, h_2 \, \zeta) \, \bar{s} s \,.
\end{eqnarray}
In the quark mean field model, quarks are confined in baryons
by the Lagrangian ${\cal L}_c=-\bar{\Psi} \, \chi_c \, \Psi$ (with $\chi_c$
given in Eq. (\ref{Dirac}), below).
The Dirac equation for a quark field $\Psi_{ij}$ under the additional
influence of the meson mean fields is given by
\begin{equation}
\left[-i\vec{\alpha}\cdot\vec{\nabla}+\chi_c(r)+\beta m_i^*\right]
\Psi_{ij}=e_i^*\Psi_{ij}, \label{Dirac}
\end{equation}
where $\vec{\alpha} = \gamma^0 \vec{\gamma}$\,, $\beta = \gamma^0$\,,
the subscripts $i$ and $j$ denote the quark $i$ ($i=u, d, s$)
in a baryon of type $j$ ($j=N, \Lambda, \Sigma, \Xi$)\,;
$\chi_c(r)$ is a confinement potential, i.e. a static potential
providing confinement of quarks by meson mean-field configurations.
The quark mass $m_i^*$ and energy $e_i^*$ are defined as
\begin{equation}
m_i^*=-g_\sigma^i\sigma - g_\zeta^i\zeta+m_{i0}
\end{equation}
and
\begin{equation}
e_i^*=e_i-g_\omega^i\omega-g_\phi^i\phi \,,
\end{equation}
where $e_i$ is the energy of the quark under the influence of
the meson mean fields. Here $m_{i0} = 0$ for $i=u,d$ (nonstrange quark)
and $m_{i0} = \Delta m_s = 29$~MeV for $i=s$ (strange quark).
Using the solution of the Dirac
equation~(\ref{Dirac}) for the quark energy $e_i^*$
it has been common to define
the effective mass of the baryon $j$ through the ansatz:
\begin{eqnarray}
M_j^*=\sqrt{E_j^{*2}- <p_{j \, cm}^{*2}>} \label{square}\,,
\end{eqnarray}
where $E_j^*=\sum_in_{ij}e_i^*+E_{j \, spin}$ is the baryon energy and
$<p_{j \, cm}^{*2}>$ is the subtraction of the contribution
to the total energy associated with spurious center of mass
motion. In the expression for the baryon energy $n_{ij}$ is the number
of quarks with flavor $"i"$ in a baryon
with flavor $j$, with $j = N \, \{p, n\}\,,
\Sigma \, \{\Sigma^\pm, \Sigma^0\}\,, \Xi \,\{\Xi^0, \Xi^-\}\,,
\Lambda\,$  and $E_{j \, spin}$ is the correction
to the baryon energy which is determined from a fit to the data for
baryon masses.

There is an alternative way to remove the spurious c.m.\ motion and
determine the effective baryon masses. In Ref.~\cite{Guichon2},
the removal of the spurious c.m.\ motion for three quarks moving in
a confining, relativistic oscillator potential was studied in some
detail. It was found that when an external scalar potential was
applied, the effective mass obtained from the interaction Lagrangian
could be written as
\begin{eqnarray}
M_j^*=\sum_in_{ij}e_i^*-E_j^0 \label{linear}\,,
\end{eqnarray}
where $E_j^0$ was found to be only very weakly dependent on the
external field strength.
We therefore use Eq.~(\ref{linear}), with $E_j^0$ a
constant, independent of the density, which is adjusted to give a
best fit to the free baryon masses.

Using the square root ansatz for the effective baryon
mass, Eq.~(\ref{square}), the confining potential
$\chi_{c}$ is chosen as a combination of scalar
(S) and scalar-vector (SV) potentials as in Ref.~\cite{Wang4}:
\begin{eqnarray}
\chi_{c}(r)=\frac12 [\,\chi_{c}^{\rm S}(r)
                         + \chi_{c}^{\rm SV}(r)\,]
\end{eqnarray}
with
\begin{eqnarray}
\chi_{c}^{\rm S}(r)=\frac14 k_{c} \, r^2 \,,
\end{eqnarray}
and
\begin{eqnarray}
\chi_{c}^{\rm SV}(r)=\frac14 k_{c} \, r^2(1+\gamma^0) \,.
\end{eqnarray}
On the other hand, using the linear definition of effective baryon
mass, Eq.~(\ref{linear}), the confining potential
$\chi_{c}$ is chosen to be the purely scalar potential $\chi_{c}^{\rm S}(r)$.
The coupling $k_{c}$ is taken as
$k_{c} = 1$ (GeV fm$^{-2})$, which yields baryon radii of about 0.6 fm.
In both cases, the baryon masses in vacuum are chosen as: $M_N=939$
MeV, $M_\Lambda=1116$ MeV, $M_\Sigma=1196$ MeV and $M_\Xi=1318$
MeV.

\bigskip
\bigskip

\section{hadronic system}

Based on the previously defined quark mean field model
the effective Lagrangian for the study of hadronic
systems is written as
\begin{eqnarray}
{\cal L}_H &=&\bar{\psi}_B(i\gamma^\mu\partial_\mu-M_B^*)\psi_B
+\frac12\partial_\mu\sigma\partial^\mu\sigma+\frac12
\partial_\mu\zeta\partial^\mu\zeta+\frac12\partial_\mu
\chi\partial^\mu\chi-\frac14F_{\mu\nu}F^{\mu\nu}
-\frac14S_{\mu\nu}S^{\mu\nu} \nonumber \\
&&-\frac14E_{\mu\nu}E^{\mu\nu}
-g_\omega^B\bar{\psi}_B\gamma_\mu\psi_B\omega^\mu
-g_\phi^B\bar{\psi}_B\gamma_\mu\psi_B\phi^\mu
-\frac12g_\rho^B\bar{\psi}_B\gamma_\mu\psi_B\rho^\mu
\nonumber \\
&&-e_B\bar{\psi}_B\gamma_\mu \psi_BA^{\mu}
+{\cal L}_M,
\end{eqnarray}
where
\begin{equation}
F_{\mu\nu}=\partial_\mu\omega_\nu-\partial_\nu\omega_\mu,
\end{equation}
\begin{equation}
S_{\mu\nu}=\partial_\mu\phi_\nu-\partial_\nu\phi_\mu,
\end{equation}
\begin{equation}
E_{\mu\nu}=\partial_\mu A_\nu-\partial_\nu A_\mu.
\end{equation}
$e_B$ is the coupling constant of comloub interaction. The mesonic
Lagrangian
\begin{equation}
{\cal L}_M = {\cal L}_{\Sigma\Sigma} + {\cal L}_{VV}
+ {\cal L}_{\chi SB}
\end{equation}
describes the interaction between mesons which
includes the scalar meson self-interaction ${\cal L}_{\Sigma\Sigma}$,
the vector meson self-interaction ${\cal L}_{VV}$ and the explicit
chiral symmetry breaking term ${\cal L}_{\chi SB}$ defined
previously in Eqs.~(\ref{scalar}), (\ref{vector}) and (\ref{L_SB}).
The Lagrangian ${\cal L}_M$ involves scalar ($\sigma$, $\zeta$
and $\chi$) and vector ($\omega$ and $\phi$) mesons.
The interactions
between quarks and scalar mesons result in the effective baryon masses
$M_B^*$, where subscript $B$ labels the baryon flavor
$B = N, \Lambda, \Sigma$ or $\Xi$.
The interactions between quarks and vector mesons generate the
baryon-vector meson interaction terms. The
corresponding vector coupling constants $g_\omega^B$ and $g_\phi^B$
satisfy the $SU(3)$ flavor symmetry relations:
\begin{equation}
g_\omega^\Lambda=g_\omega^\Sigma=2g_\omega^\Xi=\frac23g_\omega^N
=2g_\omega^u=\frac{g_v}{\sqrt{2}}
\hspace*{.5cm} \mbox{and} \hspace*{.5cm}
g_\phi^\Lambda=g_\phi^\Sigma=\frac12g_\phi^\Xi=\frac{\sqrt{2}}
3g_\omega^N=g_\phi^s=\frac{g_v}{2} .
\end{equation}
Compared with the chiral $SU(3)$ model of Refs. \cite{Papazoglou1}
and \cite{Papazoglou2}, the mesonic Lagrangian is the same.
However, that model is based on the hadron degree of freedom and
the effective baryon masses are obtained
from the direct baryon-meson coupling. This chiral $SU(3)$ quark mean
field model is on the quark level and the effective baryon masses are
generated by quarks which couple with mesons and are
confined in baryons by a potential. In the earlier work, the
effective baryon masses are calculated with the square root ansatz.
In this paper, we use the improved linear definition of baryon masses.

The equations for mesons $\phi_{i}$ can be obtained by
the formula $-\frac{\partial{\cal L}_H}{\partial^\mu(\partial_\mu\phi)}
-\frac{\partial{\cal L}_H}{\partial\phi}=0$.
Therefore, the equations for $\sigma$, $\zeta$ and $\chi$ are
\begin{eqnarray}
-\partial_\mu\partial^\mu\sigma+k_0\chi^2\sigma-4k_1\left(\sigma^2+\zeta^2
\right)\sigma-2k_2\sigma^3
-2k_3\chi\sigma\zeta-\frac{2\delta}{3\sigma}\chi^4+\frac{\chi^2}{\chi_0^2}
m_\pi^2F_\pi \nonumber \\
-\left(\frac{\chi}{\chi_0}\right)^2m_\omega\omega^2\frac{\partial m_\omega}
{\partial\sigma}+\sum_{B = N\,, \Lambda\,, \Sigma\,, \Xi }
\frac{\partial M_{B}^{\ast }}{\partial \sigma } <\bar{\psi _{B}}\psi_{B}>=0,~~~~~~~~~~
\end{eqnarray}
\begin{eqnarray}
-\partial_\mu\partial^\mu\zeta+k_0\chi^2\zeta-4k_1\left(\sigma^2+\zeta^2\right)
\zeta-4k_2\zeta^3
-k_3\chi\sigma^2-\frac{\delta}{3\zeta}\chi^4+\frac{\chi^2}{\chi_0^2}
\left(\sqrt{2}m_k^2F_k-\frac1{\sqrt{2}}m_\pi^2F_\pi\right) \nonumber \\
-\left(\frac{\chi}{\chi_0}\right)^2m_\phi\phi^2\frac{\partial m_\phi}
{\partial\zeta}+\sum_{B = \Lambda\,, \Sigma\,, \Xi}
                \frac{\partial M_{B}^{\ast}}{\partial\zeta}
 <\bar{\psi_{B}}\psi_{B}>=0,~~~~~~~~~~~~~~~~~~~~~~~
\end{eqnarray}
\begin{eqnarray}
-\partial_\mu\partial^\mu\chi+k_0\chi\left(\sigma^2+\zeta^2\right)-k_3\sigma^2
\zeta+\left(4k_4+1
+4\ln \frac{\chi}{\chi_0}-\frac{4\delta}{3}\ln \frac{\sigma^2\zeta}{\sigma_0^2
\zeta_0}
\right)\chi^3 \nonumber \\
+\frac{2\chi}{\chi_0^2}\left[m_\pi^2F_\pi\sigma+
\left(\sqrt{2}m_k^2F_k-\frac1{\sqrt{2}}m_\pi^2F_\pi\right)\zeta\right]
-\frac{\chi}{\chi_0^2}m_\omega^2\omega^2=0.
\end{eqnarray}
For infinite hadronic matter, the meson mean field is independent
of position and $<\bar{\psi}_B \psi_B>$ is expressed as
\begin{eqnarray}
<\bar{\psi}_B \psi_B>&=&\frac{g_B \, M_B^\ast}{\pi^2} \,
\int_{0}^{k_{F_B}} dk \frac{k^2}{\sqrt{M_B^{\ast 2}+k^2}} \\
&=& \frac{g_B \, M_B^{\ast 3}}{2 \, \pi^2} \,
\biggl[ \frac{k_{F_B}}{M_B^\ast} \,
\sqrt{1 + \frac{k_{F_B}^2}{M_B^{\ast 2}}}
 -  \ln \biggl( \frac{k_{F_B}}{M_B^\ast} +
\sqrt{1 + \frac{k_{F_B}^2}{M_B^{\ast 2}}} \biggr) \biggr].  \nonumber
\end{eqnarray}
{}For finite nuclei, the integration in $<\bar{\psi_B}\psi_B>$ will change to
a sum over the states $\psi_B^\alpha$, which are obtained
from the Dirac equation for the baryon $B$ in state $\alpha$
\begin{equation}
\left[-i\gamma\cdot\vec{\nabla}+M_B^*+g_\omega^B\omega
\gamma_0+g_\phi^B\phi\gamma_0+\frac12g_\rho^B\rho\gamma_0
+e_BA_0\gamma_0\right]\psi_\alpha^B=
\epsilon_\alpha\gamma_0\psi_\alpha^B.
\end{equation}

This set of coupled equations is solved iteratively.
At each iteration, we first solve the equations for the quark wave
function using fourth-order Runge-Kutta for given meson
fields and obtain the effective baryon masses. We then solve
the equations for the nucleon radial wave functions with the same
method. The corresponding scalar and vector densities can be
obtained. Then the mean meson fields which will be used
in the next iteration step are determined for the calculated densities.
The iteration is stopped when convergence is achieved.
The energy for the finite system within the mean field approximation
can be derived in the standard way.
\begin{equation}
E=\sum_{j=N,\Lambda,\Xi;\alpha=1}^\Omega\epsilon_\alpha^j
(2j_\alpha+1)-\frac12\int d^3r((\sigma\frac{\partial M_j^*}{\partial\sigma}
+\zeta\frac{\partial M_j^*}{\partial\zeta})\rho_s^j+g_\omega^j
\omega\rho_v^j+g_\phi^j\phi\rho_v^j+g_\rho^j\rho\rho_v^j)
+E_{rearr},
\end{equation}
where $\rho_s^j=\bar{\psi}_j\psi_j$ and $\rho_v^j=\bar{\psi}_j\gamma_0\psi_j$.
$\epsilon_\alpha^j$ are the
Dirac single particle energies and $j_\alpha$ are the total
angular momenta of the single particle states. By using the equations of
meson fields, the rearrangement energy $E_{rearr}$ can be written as
\begin{eqnarray}
E_{rearr}&=&\int d^3r\left\{g_4(\omega^4+\rho^4+6\omega^2\rho^2
+2\phi^4)-k_1(\sigma^2+\zeta^2)^2-k_2\left(\frac{\sigma^4}2
+\zeta^4\right)\right. \nonumber \\
&&\left. -\frac12k_3\chi\sigma^3\zeta
+\frac{\delta}3
\chi^4\ln\left(\frac{\sigma^2\zeta}{\sigma_0^2\zeta_0^2}
\right)+\frac12\frac{\chi^2}{\chi_0^2}m_\pi^2f_\pi\sigma
\right. \nonumber \\
&&\left.+\frac12\frac{\chi^2}{\chi_0^2}\left(\sqrt{2}
m_K^2f_K-\frac{1}{\sqrt{2}}m_\pi^2f_\pi\right)\zeta\right\}-V_{vac},
\end{eqnarray}
where the constant $V_{vac}$ is the vacuum energy which is
subtracted to yield zero energy in the vacuum.

\bigskip
\bigskip

\section{numerical results}

Before doing detailed numerical studies, we first determine the
parameters in the model. The coupling constant $g_s$ as well as $\Delta m_{s0}$
are determined by the constituent quark masses, $m_q$ and $m_s$. $m_v$
and $\mu$ are obtained by fitting the vector meson masses. The
confining coefficient, $k_c$, is chosen to be 1000 MeVfm$^{-2}$ to
make the baryon radii (in the absence of a pion cloud
\cite{Hackett-Jones:2000js}) around 0.6 fm. For
nuclear matter, there are seven other
parameters, $k_0$, $k_1$, $k_2$, $k_3$, $k_4$, $g_4$ and $g_v$,
to be determined. We fit them to the $\pi$-meson mass,
$K$-meson mass and the average mass of $\eta$ and $\eta^\prime$
which are given by the eigenvalues of the mass matrix
\begin{equation}
M_{ij}=-\frac{\delta^2{\cal L}_H}{\delta\phi_i\delta\phi_j}.
\end{equation}
There are two constraints associated with the saturation properties of nuclear
matter. We choose the parameters to fit the binding energy of
nuclear matter $\varepsilon/\rho-M_N=-16$ MeV at saturation density
$\rho_0=0.16$ fm$^{-3}$. The parameters should also produce a
reasonable compression modulus and effective nucleon mass at
saturation density. For strange matter, there are two additional
parameters $h_1$ and $h_2$ to be determined. They are restricted
by the hyperon potentials in hadronic matter. All these nine parameters
are listed in Table I with the two methods for computing
the effective baryon mass.
The properties of nuclear matter and the masses of $\sigma$ and
$\zeta$ are listed in the Table II. From the table, one can see
that the main difference between the two versions is that in the linear
definition (Eq.~(\ref{linear})),
the effective mass is larger. In fact, if we choose a larger value of
$g_4$, the effective mass will be smaller. However, in that case
the results for finite nuclei will not be so good.

\begin{table}
\caption{Parameters of the model.}
\begin{center}
\begin{tabular}{||c|c|c|c|c|c|c|c|c|c|c||}
version & $k_0$ & $k_1$ & $k_2$ & $k_3$ & $k_4$ & $g_s$ & $g_v$
& $g_4$ & $h_1$ & $h_2$\\
\hline
square & 4.21 & 2.26 & -10.16 & -4.38 & -0.13   & 4.76 & 10.99 & 7.5   & -2.07
& 2.90 \\
linear & 3.97 & 2.18 & -10.16 & -4.15 & -0.14 & 4.76 & 8.70 & 15.0  & -2.66
& 2.45 \\
\end{tabular}
\end{center}
\end{table}

\begin{table}
\caption{Nuclear properties and scalar meson masses.}
\begin{center}
\begin{tabular}{||c|c|c|c|c|c|c||}
version & $\rho_0$ (fm$^{-3}$) & E/A (MeV) & $M^*_N/M_N$ & K (MeV)&
$m_\sigma$ (MeV) & $m_\zeta$ (MeV) \\ \hline
square & 0.16 & -16.0 & 0.603 & 225  & 466.2 & 1167.1  \\
linear & 0.16 & -16.0 & 0.742 & 303  & 487.8 & 1168.0  \\
\end{tabular}
\end{center}
\end{table}

\subsection{Infinite hadronic matter}

We first consider infinite nuclear matter. In Fig. 1, we plot
the effective nucleon mass versus meson mean field $\sigma$. The
solid and dashed lines correspond to the linear (\ref{linear})
and square root (\ref{square}) definitions of baryon mass,
respectively. In
vacuum, we have $\sigma_0=F_\pi\simeq 0.47$ fm$^{-1}$.
With increasing density the value of $\sigma$
increases, resulting in a decreasing effective nucleon mass.
For $\sigma>-0.2$ fm$^{-1}$, the dashed line decreases very fast,
while the slope of the solid line changes a little. In Fig. 2, we
show the effective nucleon mass divided by the free nucleon mass,
$M_N^*/M_N$, versus nuclear
density. The effective mass for the square root ansatz decreases faster
than that for the linear definition and, as a consequence,
at some critical density the effective
mass drops to zero. However, in the case of the linear definition,
the effective mass decreases slowly at high density and there is
no phase transition to a state of chiral symmetry restoration. We plot the
energy per nucleon versus nuclear density in Fig. 3. The behavior of the
solid line and dashed lines are close when the density is small, say
$\rho_B<0.2$ fm$^{-3}$. Both curves pass through the saturation point of
nuclear matter. For the square root ansatz for effective mass, E/A changes
discontinuously at the critical density.

\begin{center}
\begin{figure}[hbt]
\includegraphics[scale=0.66]{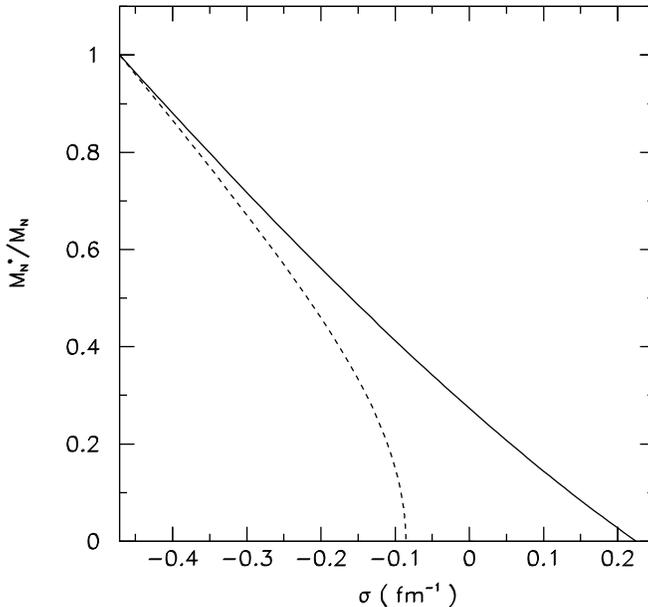}
\caption{The effective nucleon mass $M_N^*/M_N$ versus the $\sigma$ mean field
The solid and dashed lines are for linear and square root treatments of the
effective baryon mass, respectively.}
\end{figure}

\begin{figure}[hbt]
\includegraphics[scale=0.66]{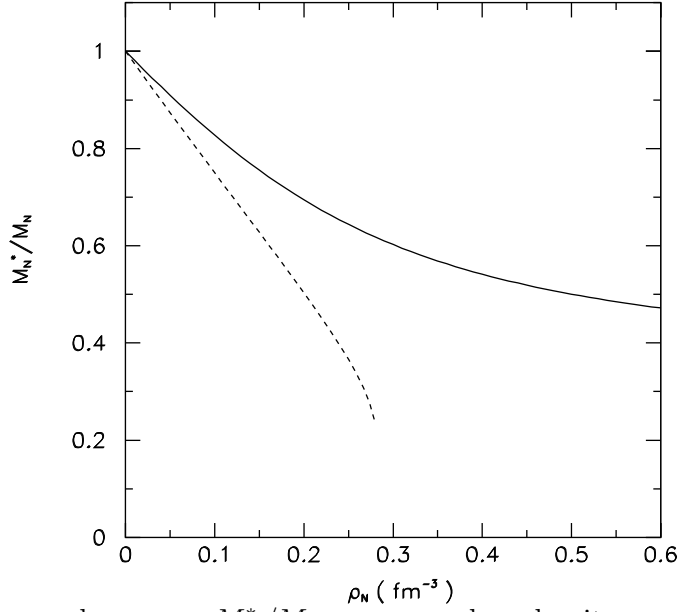}
\caption{
The effective nucleon mass $M_N^*/M_N$ versus nuclear density $\rho_N$.
The solid and dashed lines are for linear and
square root treatments of effective baryon mass, respectively.
}
\end{figure}

\begin{figure}[hbt]
\includegraphics[scale=0.66]{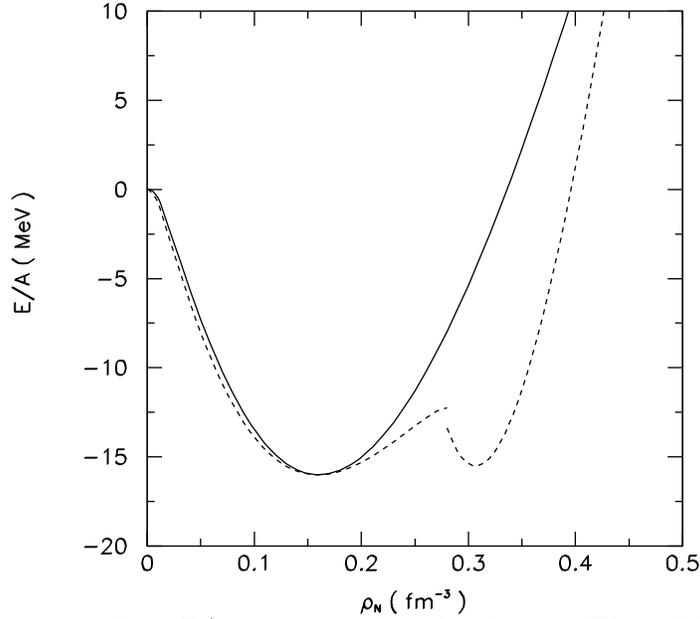}
\caption{
The energy per nucleon E/A versus nuclear density $\rho_N$.
The solid and dashed lines are for linear and
square root treatments of effective baryon mass, respectively.
}
\end{figure}
\end{center}

%\newpage
\begin{table}
\caption{Hyperon potentials in MeV.}
\begin{center}
\begin{tabular}{||c|c|c|c|c|c|c|c|c|c|c||}
version & $U^{(N)}_N$ &$U^{(N)}_\Lambda$ &$U^{(N)}_\Sigma$ &$U^{(N)}_\Xi$ &
$U^{(\Lambda)}_\Lambda$ &$U^{(\Lambda)}_\Sigma$ &$U^{(\Lambda)}_\Xi$ &
$U^{(\Xi)}_\Lambda$ &$U^{(\Xi)}_\Sigma$ &$U^{(\Xi)}_\Xi$ \\ \hline
square & -74.1 & -28.5 & -22.8 & -14.3 & -31.7  & -29.3 & -33.2 & -41.8 & -36.9
& -40.1 \\
linear & -64.0 & -28.0 & -28.0 & 8.0 & -24.5 & -24.5 & -20.6 & -30.6 & -30.6
& -50.3 \\
\end{tabular}
\end{center}
\end{table}

\begin{center}
\begin{figure}[hbt]
\includegraphics[scale=0.66]{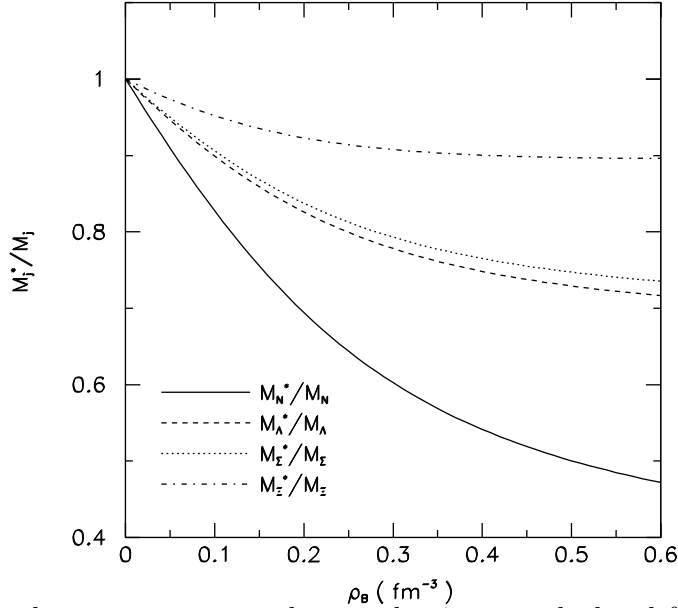}
\caption{
The effective baryon mass versus baryon density $\rho_B$
calculated for the linear definition of baryon mass with strangeness
fraction $f_s=0$. The solid, dashed, dotted and dash-dotted lines are
for nucleon, $\Lambda$, $\Sigma$ and $\Xi$, respectively.
}
\end{figure}
\end{center}

Now we consider strange hadronic matter which includes $\Lambda$,
$\Sigma$ and $\Xi$ hyperons. Since we have studied strange
hadronic matter with the square root ansatz for the baryon mass in previous
work \cite{Wang3}, we now concentrate on the linear definition
and compare the results of these two cases.
Before studying strange matter, one should first reproduce
the hyperon potentials in hadronic matter. The hyperon potential felt by
baryon $j$ in $i$-matter is defined as
\begin{equation}
U_j^{(i)}=M_j^*-M_j+g_\omega^j\omega+g_\phi^j\phi. \label{hpotential}
\end{equation}
In Table III, we list the hyperon potentials. For the $\Lambda$
hyperon, the empirical value of $U_\Lambda^{(N)}$ at the
saturation density of nuclear matter, $\rho_0$, is $-28$ MeV
\cite{Millener,Lanskoy}. For $U_\Xi^{(N)}$, recent experiments
suggest that $U_\Xi^{(N)}$ may be $-14$ or less
\cite{Fakuda,Khaustov}. In $\Lambda$ matter, the typical values of
$U_\j^{(\Lambda)}$ ($j=\Lambda, \Xi$) are around $-20$ MeV at
density $\rho=\rho_0/2$ \cite{Schaffner}. In $\Xi$ matter,
$U_\j^{(\Xi)}$ ($j=\Lambda, \Xi$) are around $-40$ MeV at density
$\rho=\rho_0$ \cite{Schaffner}. From Table III, one can see that
in the case of the square root ansatz for baryon mass, the $\Xi$
potential in hadronic matter is more reasonable, while for the
case of linear definition, the $\Lambda$ potential is more
reasonable. For the $\Sigma$ potential in nuclear matter, the
results are consistent with the earlier analysis \cite{Dover}.
Calculations in Brueckner-Hartree-Fock method show that the
potential $U_\Sigma^{(N)}$ in symmetric nuclear matter is about
-20 MeV \cite{Vidana}. However, the modern fits to $\Sigma$ atomic
data suggest a repulsive potential \cite{Mares}. Recent experiment
also shows that a strongly repulsive $\Sigma$-nucleus potential is
required to reproduce the shape of the ($\pi^-$, $K^+$) spectra on
nuclear targets \cite{Saha}. From the equation (\ref{hpotential}),
one can see that the hyperon potentials are determined by the
decrease of the effective baryon mass and the vector interaction.
Since $\Lambda$ and $\Sigma$ have the same quark components, the
calculated potentials $U_\Lambda^{(N)}$ and $U_\Sigma^{(N)}$ in
this model are close. It is important and interesting to understand
why there is a large difference of $\Lambda$ and $\Sigma$ potentials
which eventually could be considered in the model.
The precise value of $U_\Sigma^{(N)}$ will not change the main
results of the paper because $\Sigma$ hyperons do not appear in
the strange hadronic matter untill very large baryon density.
This is because the chemical potentials of $\Lambda$ and $\Sigma$
are the same, while the mass of $\Sigma$ is 80 MeV larger than
that of $\Lambda$.

\begin{center}
\begin{figure}[hbt]
\includegraphics[scale=0.66]{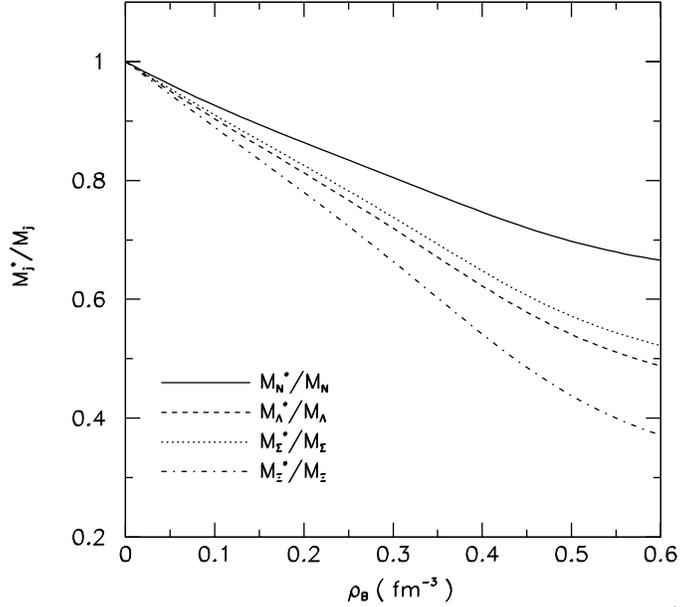}
\caption{
The effective baryon mass versus baryon density $\rho_B$
calculated for the linear definition of baryon mass with strangeness
fraction $f_s=2$. The solid, dashed, dotted and dash-dotted lines are
for nucleon, $\Lambda$, $\Sigma$ and $\Xi$, respectively.
}
\end{figure}

\begin{figure}[hbt]
\includegraphics[scale=0.66]{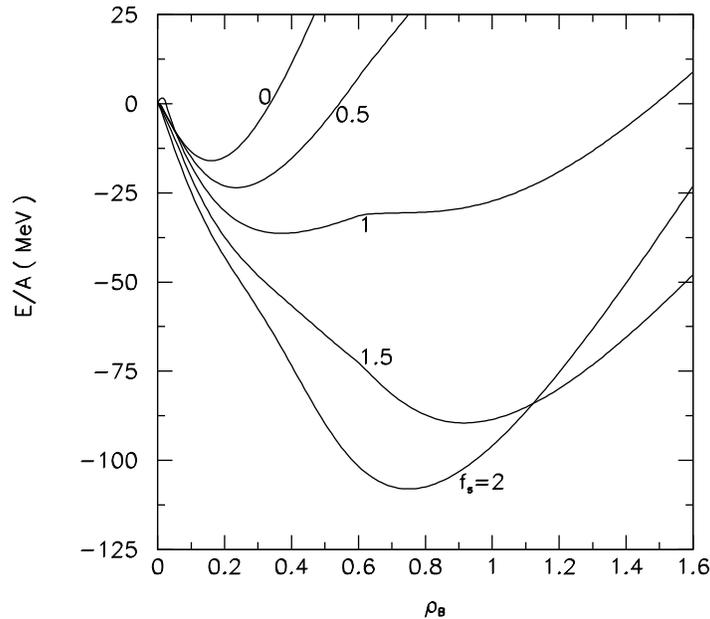}
\caption{
The energy per nucleon E/A versus baryon density $\rho_B$
calculated for the linear definition of baryon mass with differen
strangeness fraction $f_s$.
}
\end{figure}
\end{center}

In Fig. 4, we plot the effective baryon masses versus density
with strangeness fraction $f_s=0$ with the linear definition of mass.
$f_s$ is defined as
\begin{equation}
f_s=\frac{\rho_\Lambda+\rho_\Sigma+2\rho_\Xi}{\rho_B},
\end{equation}
where $\rho_\Lambda$, $\rho_\Sigma$, $\rho_\Xi$ are
the baryon densities of $\Lambda$, $\Sigma$ and $\Xi$, respectively, and
$\rho_B$ is the total density of all kinds of baryons.
All the baryon masses decrease smoothly
with increasing baryon density. The nucleon mass drops faster
than other baryons and the mass of $\Xi$ drops slowly. This is because in
nuclear matter, the interaction between nonstrange quarks is
stronger than with the strange quark. With increasing
strangeness fraction, the interaction between strange quarks
becomes more important. We show in Fig.~5 the effective
baryon mass versus density
with $f_s=2$. In contrast with Fig.~4, the mass of the $\Xi$ hyperon
drops faster than that of other baryons since the $\Xi$ has more strange
quarks. For any $f_s$, the baryon masses decrease slowly and
smoothly at high density which is different from the case of
the square root ansatz for baryon mass, where the baryon mass changes
discontinuously at some high density.
The energy per baryon versus density with different strangeness fractions
is shown in Fig.~6 for the linear definition of baryon mass.
With increasing $f_s$, the binding energy of strange hadronic
matter increases. The maximum binding energy is about 108 MeV, where
the corresponding $f_s$ is about 1.97. For the square root ansatz for
baryon mass, the largest binding energy is about 70 MeV and the
corresponding $f_s$ is about 1.5 \cite{Wang7}. The results are
comparable with those of model N in Ref. \cite{Schaffnern} where the
maximum binding energy is about 79 MeV with $f_s=1.45$.
The binding energies of various $\{N, \Lambda, \Xi\}$ and
$\{N, \Lambda, \Sigma, \Xi\}$ systems were also calculated using the
Brueckner-Hartree-Fock approximation \cite{Stoks}. Our results are
close to the case of Ref. \cite{Stoks} where the ratios of
$\Lambda$, $\Sigma$ and $\Xi$ in hadronic matter are the same.

\subsection{Finite nuclei and hypernuclei}

We now investigate the finite system. We do not adjust the
parameters, rather they are the same as in infinite hadronic
matter. The charge density versus the radius is shown in Fig. 7.
Both of the effective baryon mass definitions give reasonable
results through there is a little difference from the experiment
when the radius is smaller than 2 fm. In Fig. 8. the
charge density of $^{208}$Pb is shown. Again, the two definitions give
similar results. This arises naturally since the nucleon density is
around saturation density in the center of the finite nuclei and
both treatments yield the correct saturation properties of nuclear
matter. We plot the proton energy levels of $^{16}$O and
$^{208}$Pb in Fig. 9 and Fig. 10. The energy levels are
qualitatively reproduced. For the square root ansatz for baryon
mass, the spin-orbit splitting is quite close to the experiments.
For example, the proton spin-orbit splitting of $1p_{1/2}$ and
$1p_{3/2}$ of $^{16}$O is about 5.5 MeV which is close to the
experimental value 6.7 MeV. The splitting of $1g_{7/2}$ and
$1g_{9/2}$ ( $2d_{3/2}$ and $2d_{5/2}$ ) of $^{208}$Pb is about
4.2 MeV ( 1.6 MeV ) which is close to the experimental value 3.9
MeV ( 1.5 MeV ). For the linear definition of baryon mass, the
spin-orbit splitting is smaller. This is because at saturation
density, the effective nucleon mass is higher in this definition
and the spin-orbit splitting is proportional to the decrease of the
effective nucleon mass. However, the smaller spin-orbit
splitting can be improved by going beyond the mean field approximation
\cite{Biro}.

\begin{center}
\begin{figure}[hbt]
\includegraphics[scale=0.66]{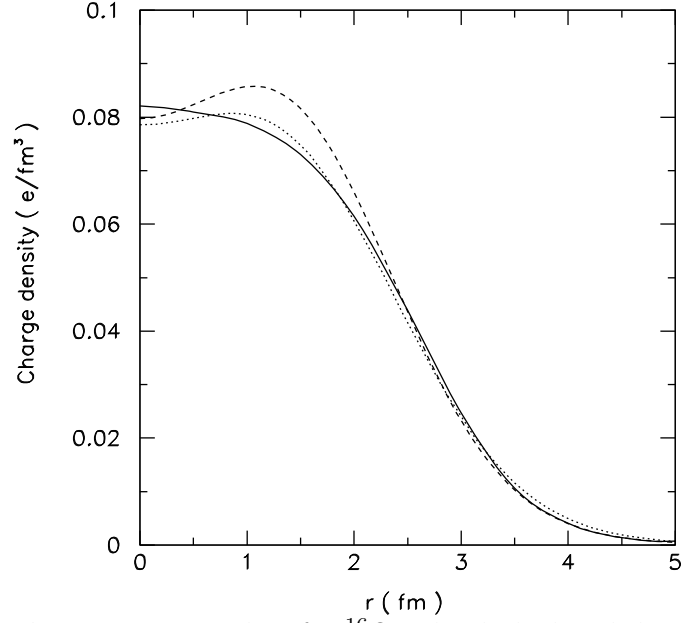}
\caption{
The charge density versus radius for $^{16}$O. The
dashed and dotted lines are calculated with linear and square root
treatments of effective baryon mass, respectively.
The solid line is from the experimental data.
}
\end{figure}

\begin{figure}[hbt]
\includegraphics[scale=0.66]{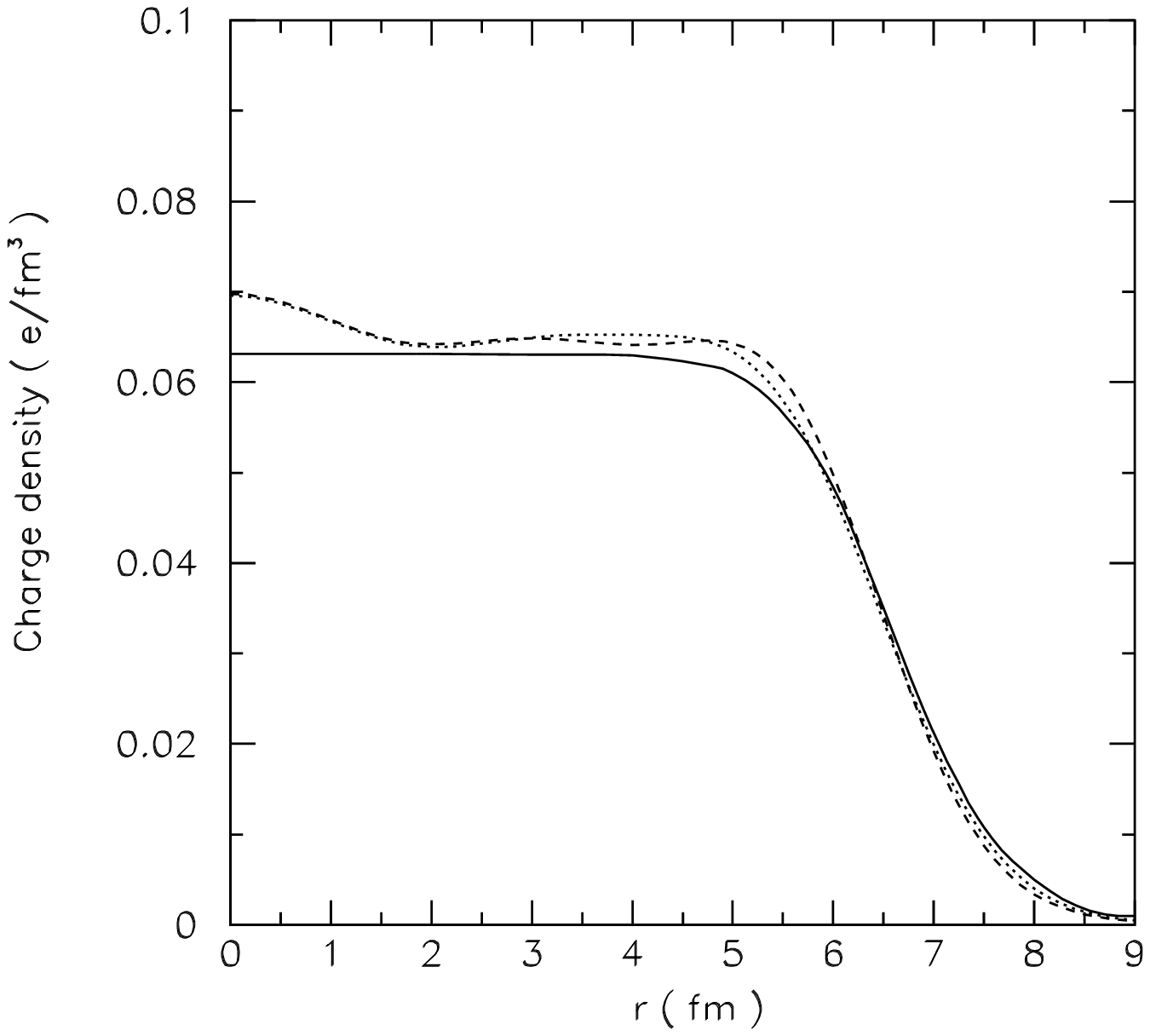}
\caption{
The charge density versus radius for $^{208}$Pb. The
dashed and dotted lines are calculated with linear and square root
treatments of effective baryon mass, respectively.
The solid line is from the experimental data.
}
\end{figure}

\begin{figure}[hbt]
\includegraphics[scale=0.66]{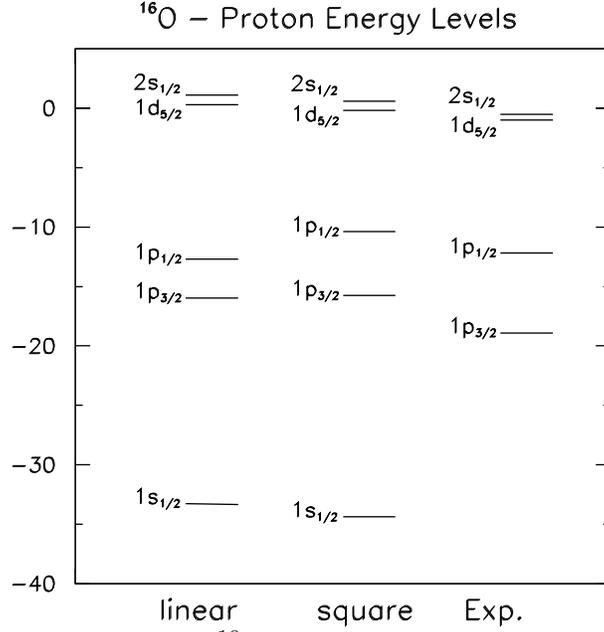}
\caption{
The proton energy levels for $^{16}$O. The first and second
columns are calculated with linear and square root treatments
of effective baryon mass, respectively. The third column is from
the experimental data.
}
\end{figure}

\begin{figure}[hbt]
\includegraphics[scale=0.66]{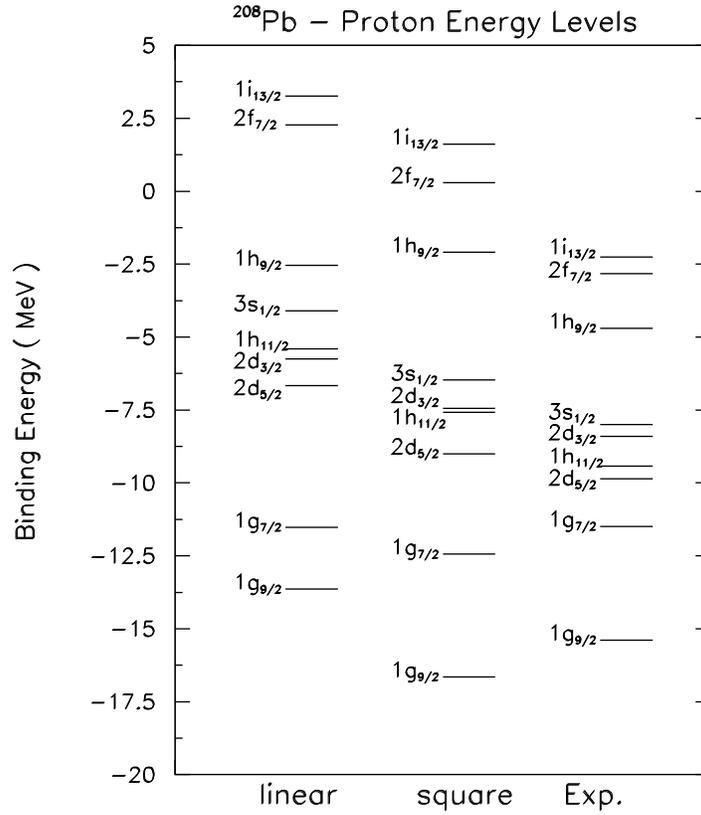}
\caption{
The proton energy levels for $^{208}$Pb. The first and second
columns are calculated with linear and square root definition
of effective baryon mass, respectively. The third column is from
the experimental data.
}
\end{figure}
\end{center}

Next we investigate hypernuclei. The hypernuclei which includes
one hyperon was studied in some detail in the QMC model \cite{Tsushima}.
We will concentrate on the binding energies of $\Lambda$
and double-$\Lambda$ hypernuclei.
The binding energy $B_\Lambda$, of $\Lambda$-hypernuclei is
expressed as
\begin{equation}
B_\Lambda=M(^{A-1}Z)+M_\Lambda-M(^A_\Lambda Z).
\end{equation}
The results are plotted in Fig.~11. The experimental values are
from Refs. \cite{Chrien,Juric}. For the square root ansatz of
baryon mass, one can see that for the light
hypernuclei, the binding energies $B_\Lambda$ are about 3 MeV
larger than the experimental values. When the baryon number is
larger than 10, the deviation from the experimental values is
around 20-30$\%$. For the heavy lambda-hypernuclei, the results
are very close to the experiment. This is because the parameters are
obtained for bulk hadronic matter and the mean field approximation
is not good when baryon number A is not large. For the linear
definition of baryon mass, the results are improved since the
$\Lambda$ potentials are more reasonable in this definition.

\begin{center}
\begin{figure}[hbt]
\includegraphics[scale=0.66]{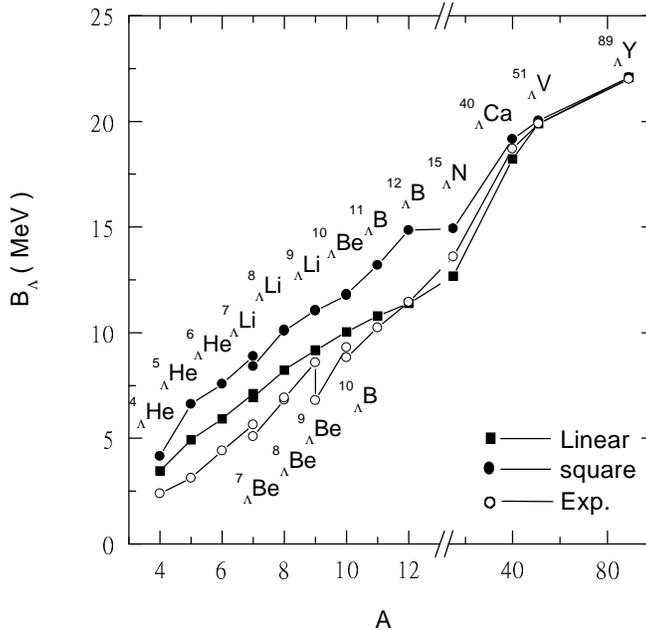}
\caption{
Hypernuclei binding energies $B_\Lambda$ calculated with linear and
square root treatments of effective baryon mass versus baryon number A.
}
\end{figure}
\end{center}

\begin{table}
\caption{The values of double-$\Lambda$ hypernuclei $B_{\Lambda\Lambda}$
and $\Delta B_{\Lambda\Lambda}$ in MeV.}
\begin{center}
\begin{tabular}{||c|c|c|c|c||c|c|c|c||}
nuclei & $B_{\Lambda\Lambda}$(L) & $B_{\Lambda\Lambda}$(S) &
Exp. \cite{Franklin}& Exp. \cite{Takahashi}& $\Delta B_{\Lambda\Lambda}$(L) &
$\Delta B_{\Lambda\Lambda}$(S)& Exp. \cite{Franklin}& Exp. \cite{Takahashi}\\
\hline
$^{~6}_{\Lambda\Lambda}$He & 11.45 & 15.21 & 10.9$\pm$0.8 & 7.25$\pm$0.19 & 1.61
& 1.97 & 4.7$\pm$1.0 & 1.01$\pm$0.20 \\
$^{10}_{\Lambda\Lambda}$Be & 19.66 & 23.66 & 17.7$\pm$0.4 & - & 1.32 & 1.60
& 4.3$\pm$0.4 & - \\
$^{13}_{\Lambda\Lambda}$B  & 23.67 & 30.22 & 27.5$\pm$0.7 & - & 0.89 & 0.54
& 4.8$\pm$0.7 & - \\
\end{tabular}
\end{center}
\end{table}

The binding energies of two lambdas $B_{\Lambda\Lambda}$
defined as
\begin{equation}
B_{\Lambda\Lambda}=M(^{A-2}Z)+2M_\Lambda-M(^A_{\Lambda\Lambda} Z)
\end{equation}
are listed in table IV.
One can see that for the square root ansatz for baryon mass,
the calculated results
are several MeV larger compared with the old experimental
values~\cite{Franklin}.
The linear definition improves this result.
There are two reports for the new events of the production and detection
of the double lambda hypernuclei \cite{Takahashi,Ahn}.
It shows that
$B_{\Lambda\Lambda}$ of $^{~6}_{\Lambda\Lambda}$He is much smaller than
the old value\cite{Takahashi}.
This new event was discussed in detail in Ref. \cite{Filikhin}.
The $\Lambda$-$\Lambda$ interaction energy $\Delta B_{\Lambda\Lambda}$ is defined
as $\Delta B_{\Lambda\Lambda}=B_{\Lambda\Lambda}-2B_\Lambda$.
The calculated results of these two treatments are similar and much
smaller than the old experimental values. $\Delta B_{\Lambda\Lambda}$ of
$^{~6}_{\Lambda\Lambda}$He obtained in this model is comparable with the new result.

\bigskip
\bigskip

\section{summary}

We have used an improved treatment of the c.m.\ motion in calculating the
effective, in-medium, baryon mass
in an investigation of infinite hadronic matter, finite nuclei and
hypernuclei within the chiral $SU(3)$ quark mean field model. The
results are compared with earlier results which used the square
root ansatz for effective mass. Both treatments fit the
saturation properties of nuclear matter and therefore, for
densities lower than the saturation density, these two treatments
give reasonably similar results. The $\Xi$ potential in hadronic matter is
reproduced better in the square root ansatz for baryon mass, while
the linear definition gives a better $\Lambda$ potential. As a
result, the binding energy of $\Lambda$ hypernuclei calculated
from the linear definition is better when compared with
experimental values. The energy levels of finite nuclei are
reasonable in both of the treatments. In the linear definition, the spin-orbit
splitting is smaller than that in the square root case. This is caused by the
different effective nucleon mass at saturation density. The
spin-orbit splitting can be improved by going beyond the mean field
approximation \cite{Biro}.

For high baryon density, the predictions of these two treatments
are quite different. There is a phase transition of chiral
symmetry restoration in the case of the square root ansatz for baryon mass.
The physical quantities, such as effective baryon mass and energy per
baryon change discontinuously at the critical density. In the linear
case, no such transition occurs. The effective baryon masses
decrease slowly at high density. The different behavior at
high density will result in significant difference for high density
physics. It is therefore of interest to study the properties of neutron
stars with these two treatments and to try to construct clearer theoretical
motivations for possible definitions of the effective baryon mass.
This will be done in the future.

\bigskip
\bigskip

\section*{Acknowledgements}
This work was supported by the Australian Research Council
and by DOE contract DE-AC05-84ER40150, under which SURA operates
Jefferson Laboratory.

\end{document}